\documentclass{aa}  

\usepackage{graphicx}
\usepackage{txfonts}

\begin{document}

   \title{Detection of the lowest mass ratio contact binary in the universe: TYC 3801-1529-1}

   \author{Kai Li
          \inst{1}\fnmsep\thanks{Corresponding author},
          Xiang Gao\inst{1}, Di-Fu Guo\inst{1}, Dong-Yang Gao\inst{1}, Xu Chen\inst{1}, Li-Heng Wang\inst{1}, Yu-Xin Xin\inst{2}, Yu-Xin Han\inst{1}, Chun-Hwey Kim\inst{3},
          \and Min-Ji Jeong\inst{3}
          }

   \institute{Shandong Key Laboratory of Optical Astronomy and Solar-Terrestrial Environment, School of Space Science and Technology, Institute of Space Sciences, Shandong University, Weihai, Shandong, 264209, China\\
              \email{kaili@sdu.edu.cn}
         \and
         Yunnan Observatories, Chinese Academy of Sciences, Kunming 650216, China\\
         \and
         Department of Astronomy and Space Science, Chungbuk National University, Cheongju 361-763, Korea\\
             }


 \authorrunning{Li et al.}
 \titlerunning{The lowest mass ratio contact binary}
  \abstract
   {This paper presents the first analysis of the contact binary TYC 3801-1529-1. We observed four sets of multiple bands complete light curves and one set of radial velocity curve of the primary component. Based on a simultaneous investigation of our observed and TESS light curves and the radial velocity curve, we found that TYC 3801-1529-1 is an extremely low-mass-ratio, medium contact binary with $q=0.0356$, with the contribution of the third light at a level of about 10\%. Its mass ratio is lower than V1187 Her, making  TYC 3801-1529-1  the lowest mass-ratio contact binary ever found in the universe. The light curves observed in 2022 are asymmetric, which is aptly explained by a hot spot on the primary component. A 16-year eclipse timings analysis indicates a secular increase orbital period with a rate of dp/dt$=7.96(\pm0.35)\times10^{-7}$ d yr$^{-1}$. We studied the stability of this target and identified that not only the value of $J_{spin}/J_{orb}$, but also the mass ratio surpass the unstable boundary. Hence, TYC 3801-1529-1 presents a challenge to theoretical research and ought to be considered a progenitor of a contact binary merger. }


   \maketitle
%

\section{Introduction}

A contact binary is one type of close binary, which captures two Roche lobe-filling component stars. The light curve of contact binary exhibits continuous variation and two nearly equivalent minima, indicating nearly equal temperatures of the two components despite their very different masses \citep{1968ApJ...151.1123L}. This type of star is very common in our Galaxy, namely, it is estimated that about 1 of every 500 stars is a contact binary \citep{2007MNRAS.382..393R}. They play a very important role in determining the physical parameters and investigating the structure and evolution of stars.
They are thought originate from short period detached binaries through angular momentum loss (AML) due to magnetic braking (e.g. \citealt{1988ASIC..241..345G,1994ASPC...56..228B,2017RAA....17...87Q}) and will ultimately coalesce into quickly rotating single stars and produce a luminous red nova event \citep{2006Ap&SS.304...25Q,2011A&A...528A.114T}. Statistical studies prove that the presence of additional companions is very common for short period binaries, especially contact binaries (e.g. \citealt{2006A&A...450..681T,2010MNRAS.405.1930L}). Therefore, additional companions are thought to paly an important role during the formation of contact binaries by removing angular momentum due to Kozai-Lidov mechanism and tidal friction (e.g. \citealt{1962AJ.....67..591K,1962P&SS....9..719L,2007ApJ...669.1298F,2019MNRAS.485.4588L}).

To study the formation, evolution, and ultimate fate of contact binaries, their accurate physical parameters should be acquired. The radial velocity curve and light curve are essential for determining accurate physical parameters of contact binaries. However, it is very difficult to derive the radial velocities of both the two components of extremely low-mass-ratio contact binaries because of the overly faint, less massive secondary and the broadened and blended spectral lines \citep{2019AJ....158..186K}. For such systems, we can only obtain accurate physical parameters for those showing flat-bottom minimum. Thanks to the statistical study of \cite{2003CoSka..33...38P}, \cite{2021AJ....162...13L}, and \cite{2021ApJS..254...10L}, contact binaries with flat-bottom minimum can help in determining accurate photometric mass ratios. The numerical simulations carried out by \cite{2005Ap&SS.296..221T} further confirmed this conclusion.

Many theoretical studies have proposed that contact binaries have a cut-off mass ratio. By neglecting the spin angular momentum of the secondaries of contact binaries, \cite{1995ApJ...444L..41R} derived a minimum mass ratio of about $q_{min}\sim0.09$. By considering the angular momenta of both components, \cite{2006MNRAS.369.2001L} found a cut-off mass ratio around 0.076-0.078. \cite{2007MNRAS.377.1635A,2009MNRAS.394..501A} assumed a radiative primary component and a fully convective secondary component, and determined that the minimum mass ratio is from 0.070 to 0.109. \cite{2010MNRAS.405.2485J} suggested that the structure of the primary affects the minimum mass ratio and they obtained a theoretical minimum mass ratio from 0.05 to 0.105 by considering the structure of the primary components with different masses. \cite{2015AJ....150...69Y} carried out a statistical study on low-mass-ratio, deep contact binaries and determined a cut-off mass ratio of about 0.044 given the relations of $q-f$ and $q-J_{spin}/J_{orb}$. \cite{2021MNRAS.501..229W,2024MNRAS.527....1W} found that the instability mass ratio is related to the mass of the primary,  fill-out factor, and  metallicity. By establishing the relationship between the critical gyration radius of the primary component and the mass ratio, \cite{2024NatSR..1413011Z} derived a new cut-off mass ratio for contact binaries: q$_{min}\approx0.038\sim0.041$. \cite{2024SerAJ.208....1A} have given an overview of the instability condition for contact binaries. They proposed a method for identifying potential merging progenitors. Theoretical studies have posited that contact binaries will merge into a single, rapidly rotation object when the mass ratio is lower than the cut-off mass ratio (e.g. \citealt{1995ApJ...444L..41R,2006MNRAS.369.2001L,2010MNRAS.405.2485J}). However, only one contact binary merging event, V1309 Sco, has been observed until now \citep{2011A&A...528A.114T}. Therefore, searching for and investigating contact binaries with extremely low ratios has been a hot topic of the astrophysical research (e.g. \citealt{2017PASJ...69...79L,2021ApJ...922..122L,2022AJ....164..202L,2022MNRAS.512.1244C,2022MNRAS.517.1928G,2023PASP..135i4201W,2023PASP..135g4202W,2023MNRAS.519.5760L,2024AJ....167..148C}).

TYC 3801-1529-1 (V$=12.340$ mag) was firstly catalogued as a contact binary by \cite{2020ApJS..249...18C}. Its orbital period is determined to be 0.3659090 days and the light variation amplitude in g band is 0.108 mag. Due to its such low amplitude variability, TYC 3801-1529-1 is thought to be an extremely low-mass-ratio contact binary \citep{2001AJ....122.1007R,2023A&A...672A.176P}. In this paper, we present photometric and spectroscopic observations and the first comprehensive investigation of this target.

\section{Observations}

\subsection{Photometric observations\label{sec:Photometric}}
The photometric observations of TYC 3801-1529-1 were carried out by using the Weihai Observatory 1.0 m Cassegrain telescope of Shandong University (WHOT, \citealt{2014RAA....14..719H}), the Weihai Observatory 50 cm telescope (WH50), and the 85 cm telescope at the Xinglong Station installed at the National Astronomical Observatories (XL85) from 2021 to 2024. The PIXIS 2048B CCD was installed on WHOT, resulting a field of view of $12^{'}\times12^{'}$; the Dhyana 4040BSI sCMOS camera was equipped on WH50, resulting a field of view of $36^{'}\times36^{'}$; the Andor DZ936 CCD was equipped on XL85, resulting a field of view of $32^{'}\times32^{'}$. The observation details are listed in Table \ref{tab:observation}. The IRAF\footnote{IRAF (\url{http://iraf.noao.edu/}) is distributed by the National Optical Astronomy Observatories, which are operated by the Association of Universities for Research in Astronomy, Inc., under cooperative agreement with the National Science Foundation.} package was applied to process the observed images. Firstly, the bias and flat corrections were reduced from the photometric images. Secondly, the aperture photometry method was used to determine the instrument magnitudes of the target, along with the comparison and check stars. At last, the differential photometry method was used to derive the magnitude differences between the target and the comparison and the comparison and check stars. TYC 3801-1734-1 (V$=12.335$ mag) was chosen as the comparison star and 2MASS 08415215+5647340 (V$=13.039$ mag) was chosen as the check star. Based on our observations, we used the generalised Lombe-Scargle (GLS) periodogram \citep{1982ApJ...263..835S,2009A&A...496..577Z} to determine a more precise orbital period and the result is P = 0.36591971 days.

\subsection{Spectroscopic observations\label{sec:Spectroscopic}}
The spectroscopic observations of TYC 3801-1529-1 were carried out by using the Beijing Faint Object Spectrograph and Camera (BFOSC) mounted on the 2.16 m telescope at Xinglong Observatory (XL216, \citealt{2016PASP..128k5005F}) of National Astronomical Observatories, Chinese Academy of Sciences in 2021 and 2022. The E9+G10 mode was used during the observations of XL216, the width of the slit is 1.6$^{\prime\prime}$, and the resolution per pixel is about R$\sim11000$, and the wavelength coverage is from 3300 to 10000 {\AA}. The exposure time of all the observations was set as 1800 s. The observation details are listed in Table \ref{tab:observation}. A total of 22 spectra were obtained. Each night, HD 65583 (V$=7.00$ mag) was selected as the standard star for calculating the radial velocities (RVs) of TYC 3801-1529-1. The exposure time of HD 65583 was set as 180 s. The observed images were processed using IRAF package, including the bias and flat corrections, cosmic ray removing, spectra extraction, wavelength calibration, and flux normalization. The RVs of TYC 3801-1529-1 were derived using the cross-correlation function (CCF) with HD 65583 and one Gaussian fitting to the CCF, heliocentric corrections have been made and the results are listed in Table \ref{tab:RV}.
\section{Orbital period variation} \label{sec:OBV}
\begin{figure}
\centering
\includegraphics[scale=0.3]{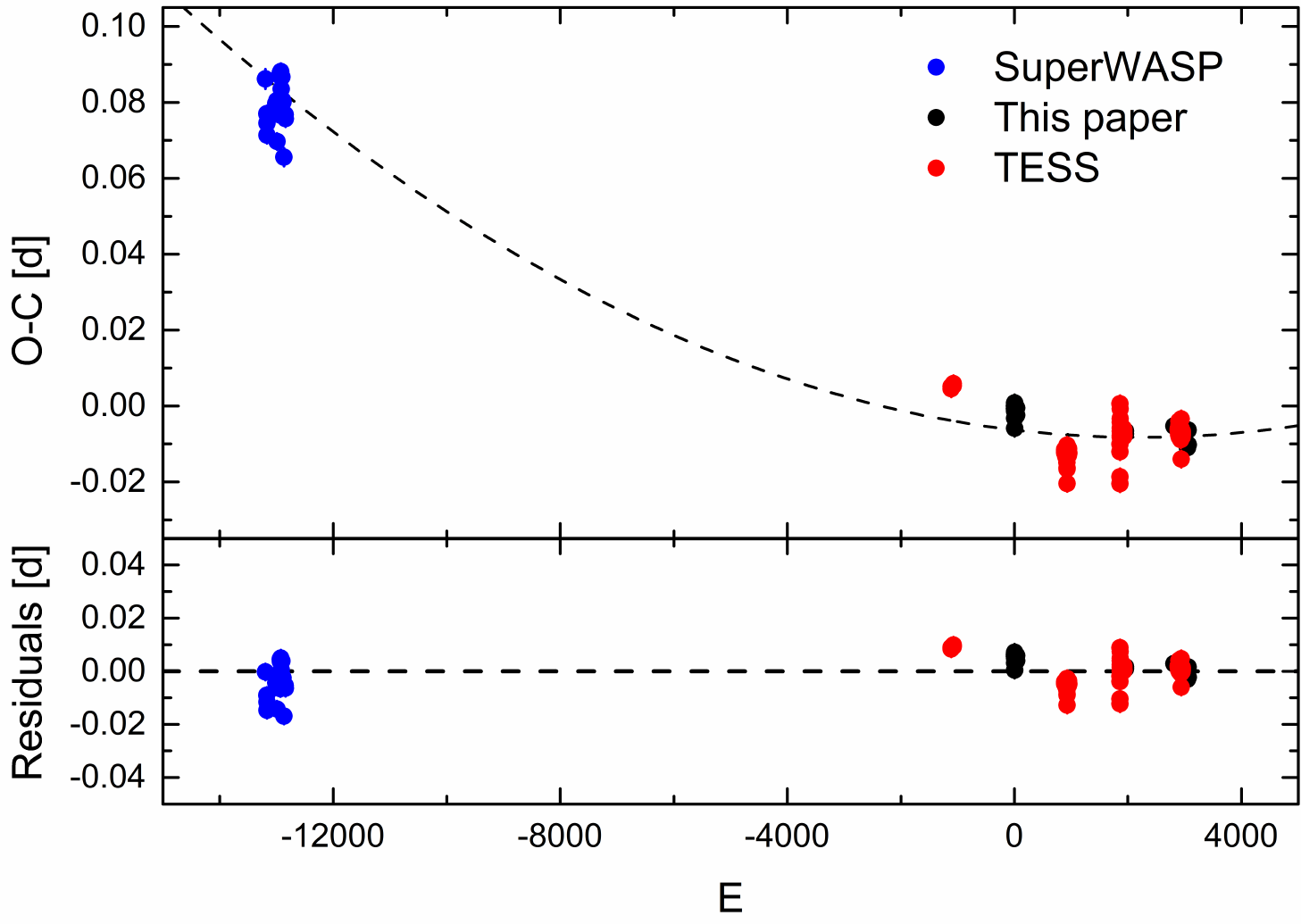}
\caption{ O-C diagram of TYC 3801-1529-1. Upper panel shows the O-C values and the parabolic fitting, lower panel displays the fitting residuals. }
\label{Fig1}
\end{figure}

The orbital period variation investigation is an important tool to analyse the dynamical evolution and search for additional companions of contact binaries \citep{2010MNRAS.405.1930L,2015NewA...41...17L,2015AJ....149..120L,2024ApJ...961...97Z}. Besides our observations, TYC 3801-1529-1 has also been observed by Super Wide Angle Search for Planets (SuperWASP, \citealt{2010A&A...520L..10B}) and Transiting Exoplanet Survey Satellite (TESS) \citep{2015JATIS...1a4003R} Sectors 20, 47, 60, 74. Using the SuperWASP data, we directly calculated 23 eclipse timings. Using the TESS Sectors 47, 60, and 74 data, we directly calculated 356 eclipse timings. For the TESS Sector 20 data, we used the period shift method proposed by \cite{2020AJ....159..189L} to derive four eclipse timings because of the low temporal resolution (the observation cadence of the data is 30 minutes). Using our own observations, 17 eclipse timings were obtained. All the eclipse timings were calculated using the method suggested by \cite{1956BAN....12..327K} and were converted to BJD$_{TDB}$ using the online procedure provided by \cite{2010PASP..122..935E}. They are shown in Table \ref{tab:ET}. A total of 400 eclipse timings that span more than 16 years have been obtained.
The observed minimum minus calculated minimum (O-C) values were computed by using the following equation,
\begin{eqnarray}
BJD=2459256.25969+0.^d36591971\times E,
\end{eqnarray}
where the initial primary minimum is observed by us, and the orbital period is determined by our observation using GLS periodogram. The results are listed in Table \ref{tab:ET} and drawn in Fig. \ref{Fig1}. In this figure, obvious upward parabola can be seen. Therefore, the following quadratic equation was used to fit the O-C values,
\begin{eqnarray}
T = 2459630.25338(\pm0.00035) + 0.36591795(\pm0.00000014) \times E\\ 
\nonumber + 3.99(\pm0.17)\times10^{-10} \times E^2.
\end{eqnarray}
The positive quadratic term meaning that the orbital period of TYC 3801-1529-1 is secular increase at a rate of dp/dt$=7.96(\pm0.35)\times10^{-7}$ d yr$^{-1}$.

\section{Light and radial velocity curves analysis} \label{sec:LC}
Although we only obtained the radial velocities of the primary component, we can still derive the reliable mass ratio because TYC 3801-1529-1 is a totally eclipsing contact binary (e.g. \citealt{2005Ap&SS.296..221T,2021AJ....162...13L}). The TESS data have very high precision, we analysed radial velocity curve and the light curves of our observations and TESS (except Sector 20 because of low temporal resolution) simultaneously. We should note that the SAP light curve of TESS was used in the following analysis. To improve the computing speed, the TESS data were binned to 1000 normal points. First, we had to determine the temperature of the primary component. We note the spectrum near phase=0.5 (phase=0.497)  observed by XL216. Then we used the University of Lyon Spectroscopic analysis Software (ULySS; \citealt{2009A&A...501.1269K}) to estimate the atmospheric parameters of the primary component. The normalised spectrum at phase=0.497 and fitted spectrum are illustrated in Fig. \ref{Fig1A}. The determined atmospheric parameters are as follows: $T_{eff}=$ $6206\pm13$ K, [Fe/H] = $0.418\pm0.010$ dex, $\log$ g = $4.193\pm0.024$ dex. Because the light curve of TYC 3801-1529-1 exhibits a flat bottom secondary minimum, the spectrum near phase=0.5 comes from the primary component, we set the temperature as the primary component. To check the accuracy of the temperature, we computed the temperature corresponding to the de-reddening color indices of (B-V)$_0$ and (J-K)$_0$, a mean value of 6351 K was determined. Meanwhile, TYC 3801-1529-1 was  observed by the Large Sky Area Multi-Object filber Spectroscopic Telescope (LAMOST; \citealt{2012RAA....12.1197C}) low-resolution spectra on 10 February 2014 and the atmospheric parameters were obtained, with a temperature of $5964\pm19$ K. Due to the results determined by color index and LAMOST spectrum, it indicates that the atmospheric parameters obtained by ULySS are reliable.
All the light curves were covert from time to phase using the equation $T=T_0+P\times E$, where $T_0$ is the primary minimum of each light curve, P is the orbital period. Then, the latest version of Wilson-Devinney (W-D) code \citep{1971ApJ...166..605W, 1979ApJ...234.1054W, 1990ApJ...356..613W} was applied to perform the light curve and radial velocity curve analysis.

\begin{figure}
\centering
\includegraphics[width=0.24\textwidth]{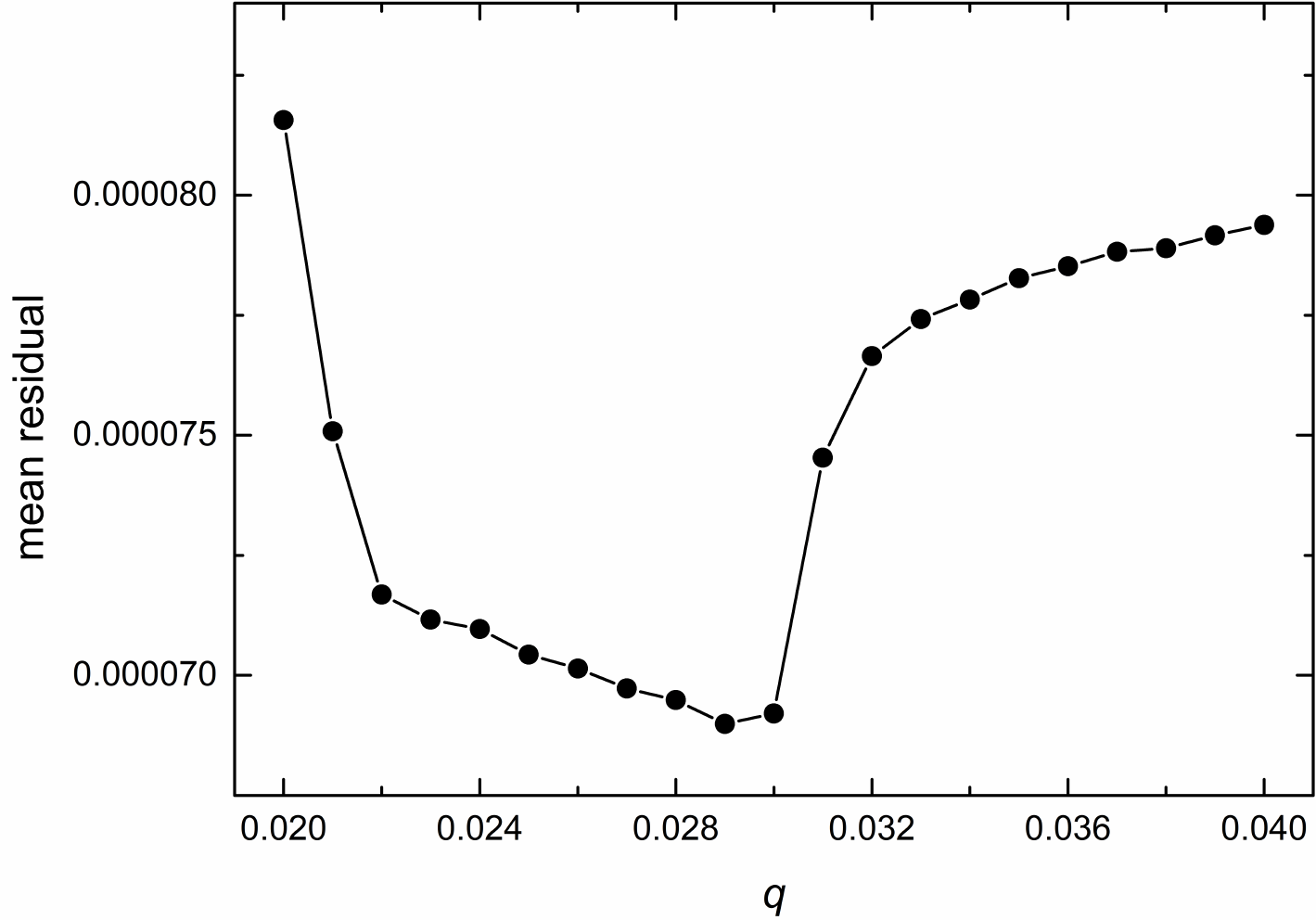}
\includegraphics[width=0.23\textwidth]{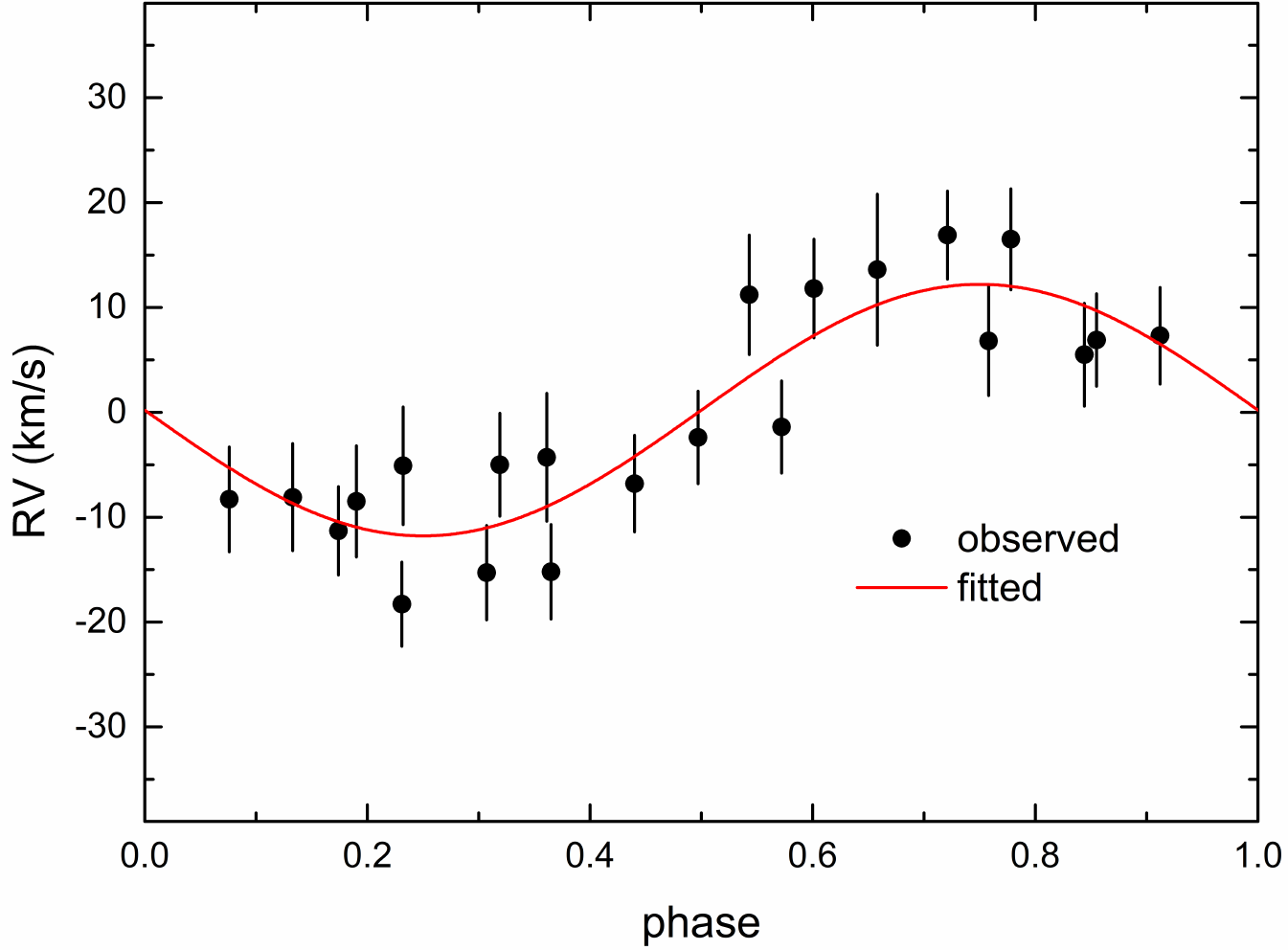}
\caption{Relationship between the mean residual and the mass ratio (left) and the RV fitting curve (right).}
\label{Fig2}
\end{figure}

\begin{figure*}[htbp]
\centering
\includegraphics[width=0.32\textwidth]{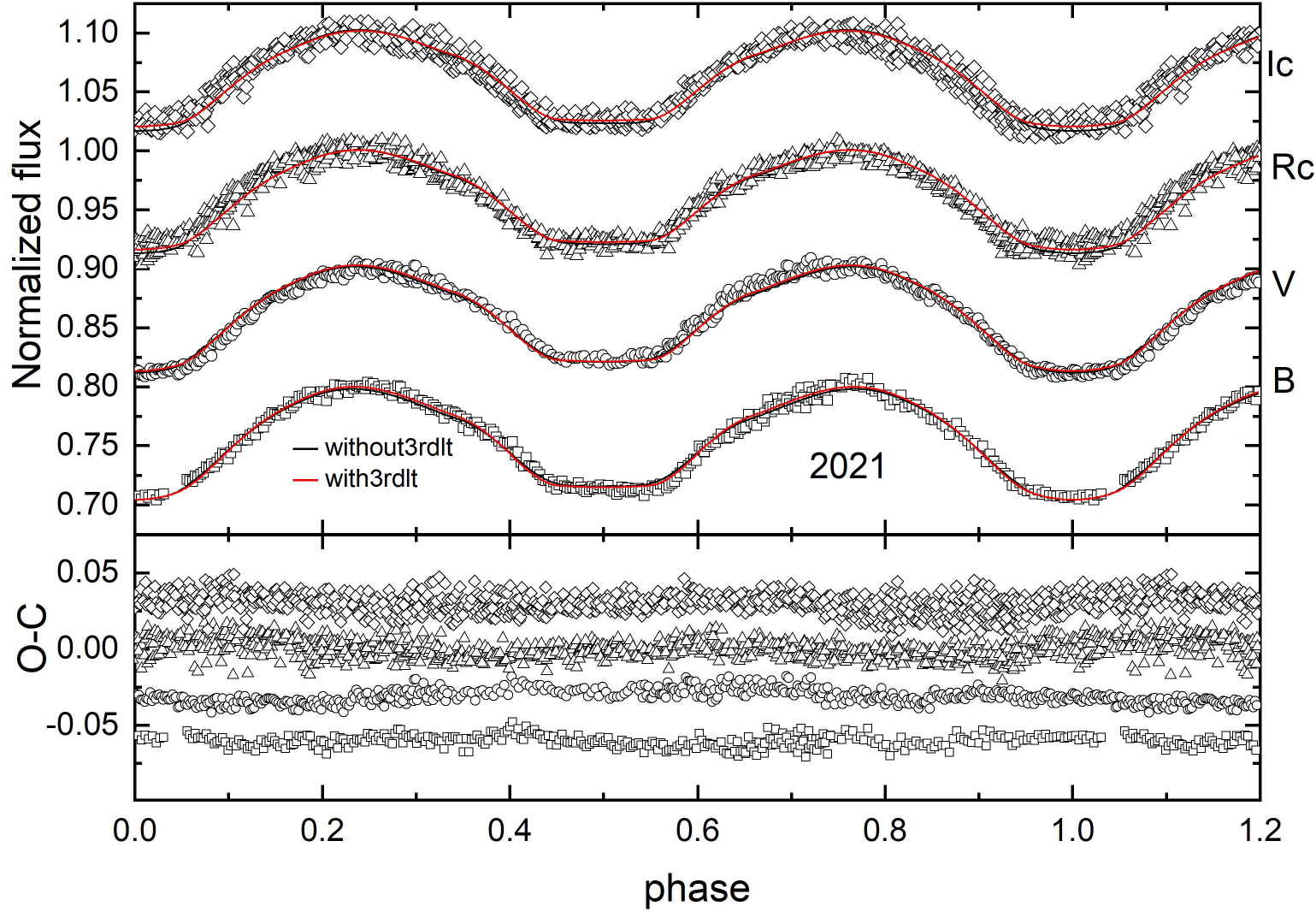}
\includegraphics[width=0.32\textwidth]{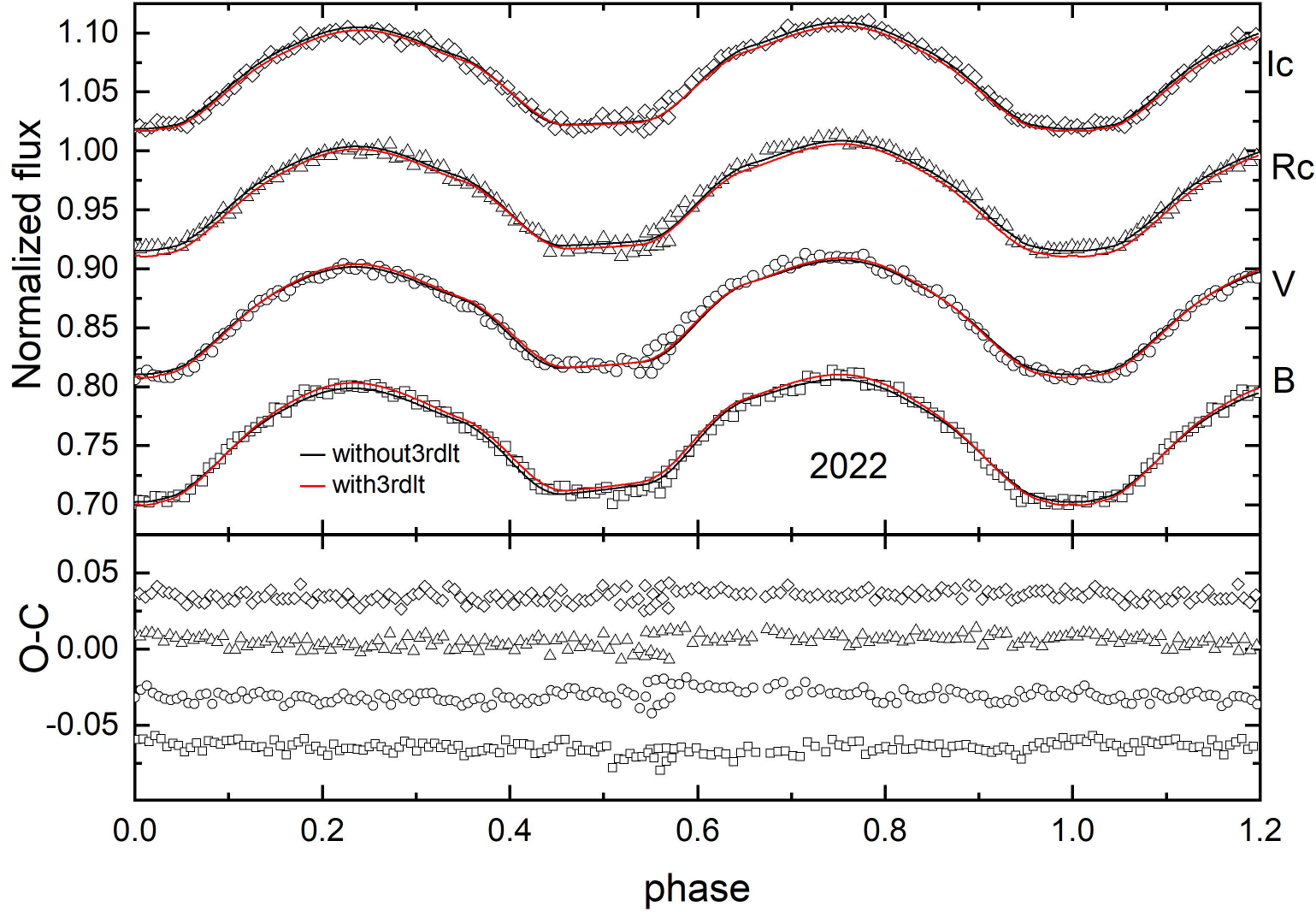}
\includegraphics[width=0.32\textwidth]{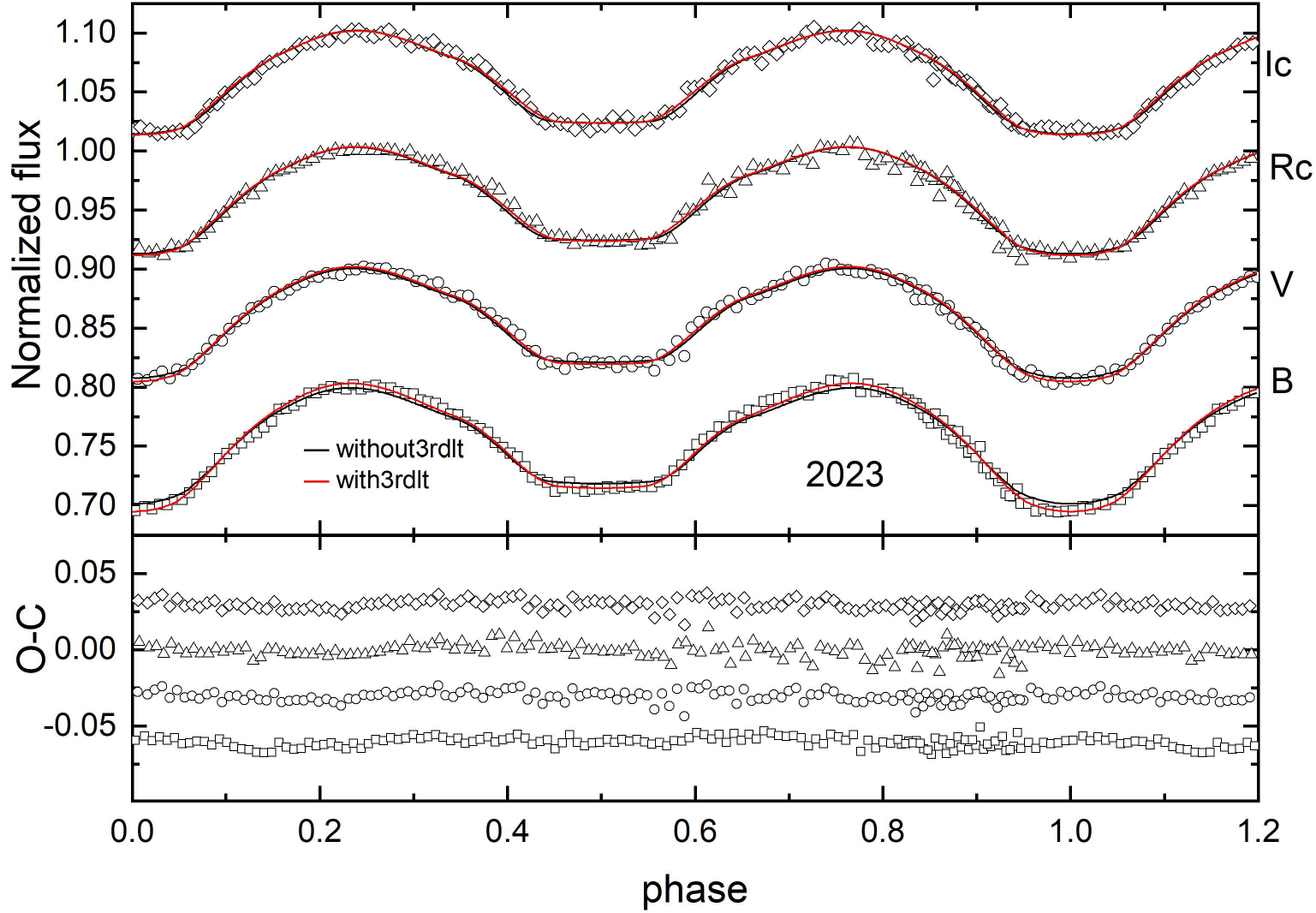}
\includegraphics[width=0.32\textwidth]{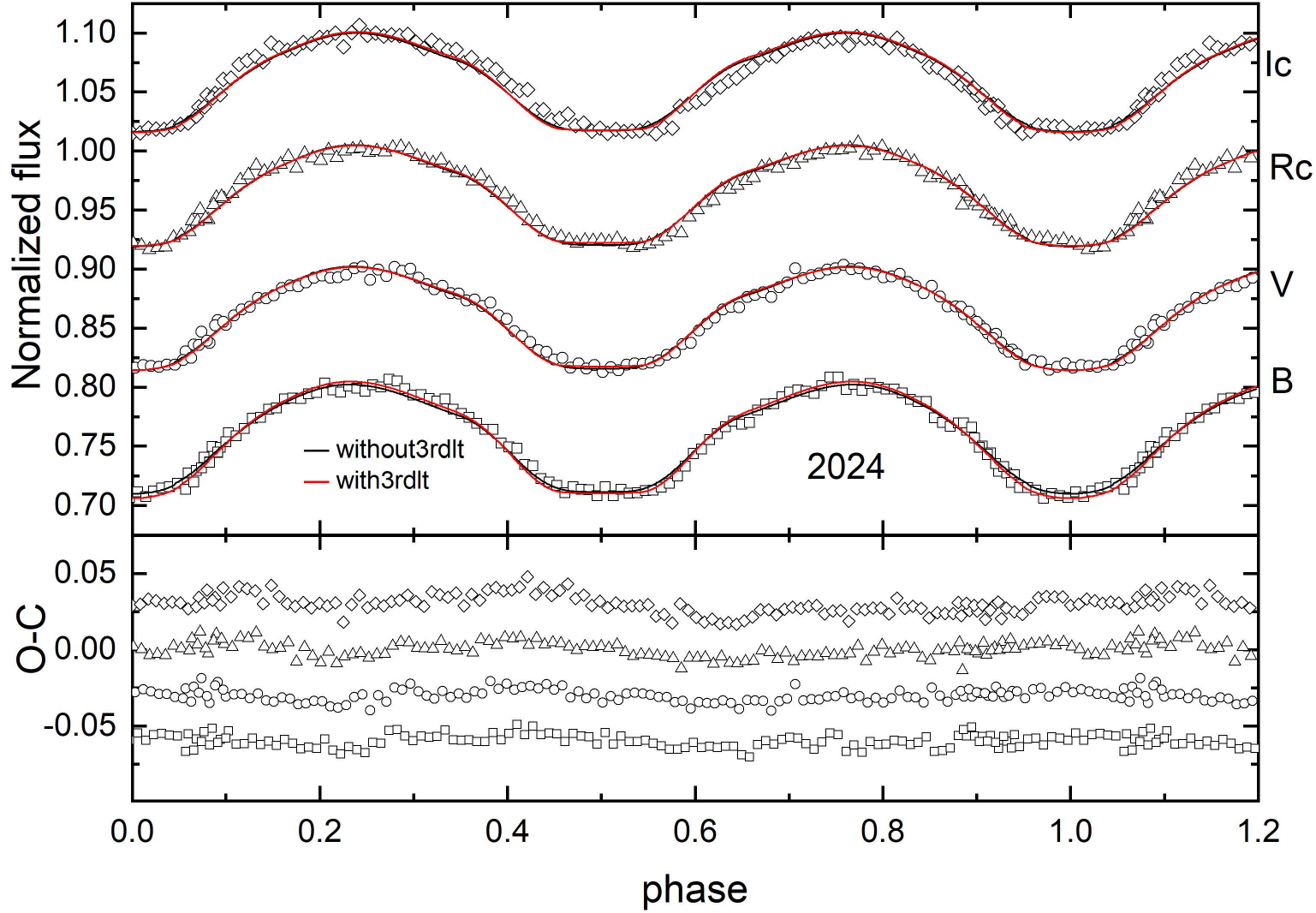}
\includegraphics[width=0.32\textwidth]{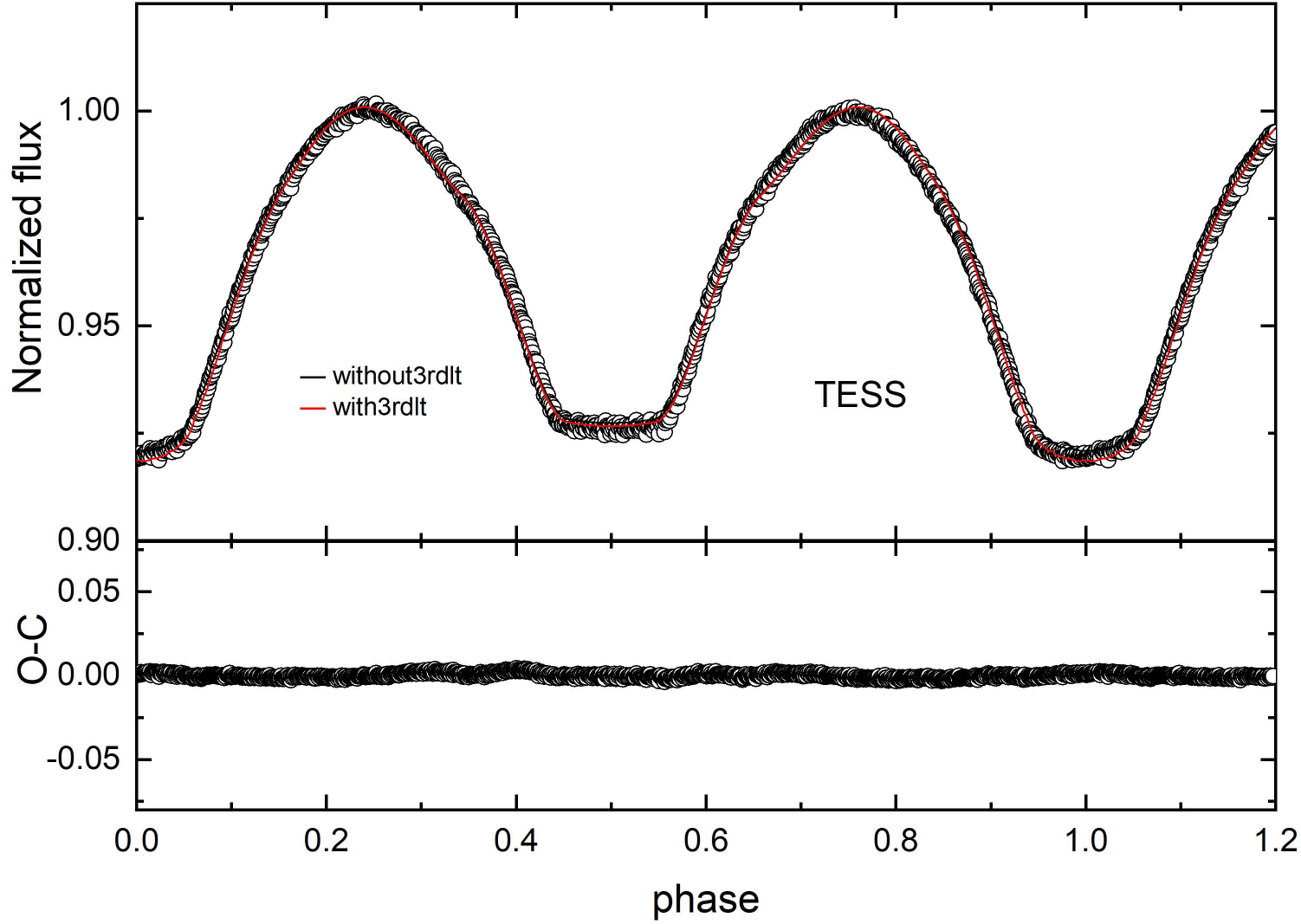}
\caption{Comparison between the theoretical light curves determined by analyzing each year light curves individually and the observed ones. Black lines show the theoretical light curves without the third light, while the red ones represent those with the third light.}
\label{Fig3}
\end{figure*}

During the modelling, the contact binary mode was chosen, and the gravity darkening and bolometric albedo coefficients were fixed at $g_1=g_2=0.32$ \citep{1967ZA.....65...89L} and $A_1=A_2=0.5$ \citep{1969AcA....19..245R}. Using the square root law (ld=-3), the limb-darkening coefficients were interpolated from the table provided by Van Hamme's personal website in 2019\footnote{\url{https://faculty.fiu.edu/~vanhamme/lcdc2015/}} \citep{1993AJ....106.2096V}. Since this binary system has not been analysed before and only the radial velocity curve of the primary has been obtained, we applied the widely used $q-$search method to determine its mass ratio. A series of solutions with fixed mass ratios were performed. The mean residual versus mass ratio, $q$, is shown in left panel of Fig. \ref{Fig2}, where we can see a clear minimum at $q=0.029$. Then, we set $q=0.029$ as an initial mass ratio and an adjustable parameter. Thus, new solutions using  the W-D code could be carried out when convergent result was determined. The light curves observed in 2022 are asymmetric and we tried to use the spot model to model the light curves. A dark spot on the secondary component or a hot spot on the primary component was tested and we found that a hot spot on the primary component leads to the optimal fitting result. We also tried to repeat the fit with third light as a free parameter and determined a convergent solution. Then, we evaluated the difference of the two fits with and  without third light using AIC and BIC criteria. AIC=-57436 and BIC=-57389 for the fit without third light, while AIC=-74043 and BIC=-73988 for the fit with third light. Statistically, the fit with third light is the better one. Therefore, we adopted the results determined with third light as the final results.
The determined physical parameters are listed in Table \ref{t:Parameters} and the theoretical radial velocity curve is shown in the right panel of Fig. \ref{Fig2}, while the theoretical light curves with and without third light are displayed in Fig. \ref{Fig2A}. From Fig. \ref{Fig2A}, we can see that some of the synthetic light curves do not fit the observed light curves very well and this is due to the simultaneous analysis of all years light curves. Therefore, we analysed each year light curves separately to gain a better fit. The comparison between the synthetic and observed light curves are shown in Fig. \ref{Fig3}, much better fitting results are obtained.

\begin{table}
\centering
\tiny
\caption{Physical parameters of TYC 3801-1529-1}
\label{t:Parameters}
\setlength{\tabcolsep}{2mm}{
\resizebox{\linewidth}{!}{
\begin{tabular}{lccccc}
\hline\hline
Parameters & 2021& 2022& 2023 & 2024&TESS \\\hline
$T_1$ (K)           &\multicolumn5c{6206}      \\
$q(M_2/M_1) $       &\multicolumn5c{0.0356$\pm0.0035$}  \\
$T_2$ (K)           &\multicolumn5c{6160$\pm241$}      \\
$i$ (deg)           &\multicolumn5c{66.7$\pm1.8$}     \\
$V_{\gamma}$ (km/s) &\multicolumn5c{0.210$\pm0.063$}     \\
$a$ (R$_\odot$)     &\multicolumn5c{2.789$\pm0.282$}  \\
$M_1$ (M$_\odot$)   &\multicolumn5c{2.096$\pm0.642$}  \\
$M_2$ (M$_\odot$)   &\multicolumn5c{0.075$\pm0.030$}  \\
$R_1$ (R$_\odot$)   &\multicolumn5c{1.831$\pm0.207$}  \\
$R_2$ (R$_\odot$)   &\multicolumn5c{0.434$\pm0.081$}  \\
$L_1$ (L$_\odot$)   &\multicolumn5c{4.455$\pm1.045$}  \\
$L_2$ (L$_\odot$)   &\multicolumn5c{0.243$\pm0.129$}  \\
$\Omega_1=\Omega_2$ &\multicolumn5c{1.724$\pm0.002$}  \\
$f$                 &\multicolumn5c{34.7$\pm8.0$\%}   \\
$L_{1B}/L_{B}$     &0.949$\pm0.022$  &0.949$\pm0.024$  &0.949$\pm0.024$  &0.949$\pm0.025$  &$-$  \\
$L_{1V}/L_{V}$     &0.949$\pm0.019$  &0.949$\pm0.022$  &0.949$\pm0.023$  &0.949$\pm0.025$  &$-$  \\
$L_{1R_c}/L_{R_c}$ &0.949$\pm0.017$  &0.949$\pm0.022$  &0.949$\pm0.023$  &0.949$\pm0.024$  &$-$  \\
$L_{1I_c}/L_{I_c}$ &0.948$\pm0.017$  &0.948$\pm0.022$  &0.948$\pm0.023$  &0.948$\pm0.023$  &$-$  \\
$L_{1T}/L_{T}$&$-$  &$-$  &$-$  &$-$  &0.948$\pm0.017$  \\
$L_{3B}/L_{B}$     &16.6$\pm1.8$\%  &7.3 $\pm2.1$\%  &9.6 $\pm2.0$\%  &14.2$\pm2.0$\%  &$-$  \\
$L_{3V}/L_{V}$     &15.9$\pm1.6$\%  &8.0 $\pm2.0$\%  &12.7$\pm2.0$\%  &19.4$\pm1.9$\%  &$-$  \\
$L_{3R_c}/L_{R_c}$ &17.5$\pm1.4$\%  &9.7 $\pm1.9$\%  &14.4$\pm1.9$\%  &19.0$\pm1.9$\%  &$-$  \\
$L_{3I_c}/L_{I_c}$ &18.0$\pm1.3$\%  &12.1$\pm1.9$\%  &14.9$\pm1.9$\%  &15.9$\pm2.0$\%  &$-$  \\
$L_{3T}/L_{T}$&$-$  &$-$  &$-$  &$-$  &18.2$\pm1.3$\%  \\
Spot                & $-$             &$-$              & star 1          &$-$              &$-$  \\
$\theta$ (deg)      & $-$             &$-$              &$92.7\pm5.5$     &$-$              &$-$  \\
$\lambda$ (deg)     & $-$             &$-$              &$146.4\pm16.2$   &$-$              &$-$  \\
$r_s$ (deg)         & $-$             &$-$              &8.8$\pm1.1$      &$-$              &$-$  \\
$T_s$               & $-$             &$-$              &1.125$\pm0.055$  &$-$              &$-$  \\
\hline
\end{tabular}}}
\end{table}

\section{Discussions and conclusions}\label{sec:discussion}
We performed the first photometric and spectroscopic study of a totally eclipsing contact binary TYC 3801-1529-1. It is found that this binary system is a very rare extremely low-mass-ratio contact binary with a mass ratio of 0.0356, the third light accounts for the total light approximately 10\%. Using Eq. (1) of \cite{2002A&A...392..895B}, we estimated the colour index of the third light to be $0.882\pm0.183$ by averaging the four years of data. If the third light is a main sequence star, it corresponds to a K2V star and the luminosity is 0.37$^{+0.42}_{-0.17}$ $L_\odot$, which is 7.3$^{+7.1}_{-3.2}$\% of the total luminosity, this is consistent with the solution of W-D code.
 V1187 Her whose mass ratio is 0.044 was considered to be the most extreme-mass-ratio contact binary \citep{2019PASP..131e4203C}; however recent studies \citep{2022AAS...24020505C,2024AAS...24342302C} suggested that very significant third light contaminates V1187 Her and a new mass ratio of 0.16 was estimated. Therefore, TYC 3801-1529-1 is found to be the lowest mass-ratio contact binary in the Universe at present. This system is an A-type W UMa contact binary with a contact degree of 35.7\%. A 16-year orbital period analysis indicates that TYC 3801-1529-1 shows a long-term increase orbital period at a rate of dp/dt$=7.96(\pm0.35)\times10^{-7}$ d yr$^{-1}$. Based on the luminosity and the third light ratio obtained by us, we can estimate the photometric distance to TYC 3801-1529-1 and determined D $=624.19\pm86.33$ pc, which is different from the result (D $=386.94\pm44.37$ pc) determined by the absolute magnitude-period-colour index calibration of \cite{1997PASP..109.1340R}.

A bright close companion star around TYC 3801-1529-1 within 0.5$^{\prime\prime}$ was identified by \textit{Gaia} DR3 \citep{2016A&A...595A...1G,2023A&A...674A...1G}. If this is true, the mass ratio of TYC 3801-1529-1 would be incorrect. Therefore, we offer a discussion  in Appendix \ref{sec:bright companion}, ruling out the possibility of the existence of this bright companion star.

\subsection*{The long-term orbital period variation}
Using 16 years times of eclipsing minima, we analysed the orbital period variation of TYC 3801-1529-1 and found that the orbital period of TYC 3801-1529-1 is secular increase with a rate of  dp/dt$=7.96(\pm0.35)\times10^{-7}$ d yr$^{-1}$. Such a change is generally caused by mass transfer from the less massive component to the more massive one. By assuming conservative mass transfer and using the following equation,
\begin{eqnarray}
{\dot{P}\over P}=-3\dot{M_1}({1\over M_1}-{1\over M_2}),
\end{eqnarray}
we can compute the mass transfer rate and the result is dm/dt$=5.60(\pm0.25)\times10^{-8}$ M$_\odot$ yr$^{-1}$. As the increasing orbital period and mass transfer, the distance between the two component will increase and the mass ratio will become more extreme. Thus, this binary would end up meeting the Darwin instability and merge into a fast ration single star.

\subsection*{The stability analysis}
The mass ratio of TYC 3801-1529-1 is only 0.0356, which is below the theoretical mass ratio limit of the contact binary \citep{1995ApJ...444L..41R,2006MNRAS.369.2001L,2007MNRAS.377.1635A,2010MNRAS.405.2485J}. Therefore, we tried to analyse the stability of this system. The spin angular momentum ($J_{spin}$) to the orbital angular momentum ($J_{orb}$) is an important examination to the dynamic stability \citep{1980A&A....92..167H}. We computed the $J_{spin}/J_{orb}$ value using the equation \citep{2006MNRAS.369.2001L}:\begin{eqnarray}
{J_{spin}\over J_{orb}}={{1+q}\over q}[(k_1r_1)^2+(k_2r_2)^2q],
\end{eqnarray}
where $k_{1,2}$ is the dimensionless gyration radius, while $r_{1,2}$ is the relative radius and $k^2_{1,2}=0.06$ was used following \cite{2006MNRAS.369.2001L}, and $J_{spin}/J_{orb}=0.755$ was obtained. The value of $J_{spin}/J_{orb}$ exceeds 1/3, thus, this binary system is expected be unstable according to the Darwin instability \citep{1980A&A....92..167H}.

\cite{2021MNRAS.501..229W} proposed a method to calculate the instability parameters, such as instability mass ratio, separation, and period. Following their method, we obtained that the instability mass ratio is $q_{inst}=0.041$, the instability separation is $A_{inst}=3.165$ R$_\odot$, and the instability period is $P_{inst}=0.361$ d. The present values of all the three parameters of TYC 3801-1529-1 are smaller than the instability parameters. Very recently, \cite{2024SerAJ.208....1A} and \cite{2024NatSR..1413011Z} also analysed the cut-off mass ratio of contact binaries and found a very similar global minimum mass ratio for different masses of contact binaries of q$_{min}=0.042\sim0.044,$ determined by \cite{2024SerAJ.208....1A}, and of q$_{min}=0.038\sim0.041,$ determined by \cite{2024NatSR..1413011Z}. The mass ratio of TYC 3801-1529-1 is below the cut-off mass ratio, however, its orbital period variation shows a normal and stable changing rate, while the light curves are also stable for the four years observations. We go on to consider why discrepancies exist between theoretical calculations and actual observations. We think this is likely to be caused by the dimensionless gyration radius. No matter whether $J_{spin}/J_{orb}$ or the instability mass ratio, both are strongly correlated with the dimensionless gyration radius. The present dimensionless gyration radius-stellar mass relation is for ZAMS stars, the gyration radius can be even smaller for evolved main sequence stars \citep{2010MNRAS.405.2485J}, then smaller $J_{spin}/J_{orb}$ and instability mass ratio can be obtained. The mass transfer and mass loss can also affect calculation of the dimensionless gyration radius.

In conclusion, based on photometric light curves and radial velocity curve investigation, we found that TYC 3801-1529-1 is an extremely low-mass-ratio medium contact binary. This system is the lowest mass ratio contact binary in the universe. Such a low-mass-ratio target poses a challenge to theoretical research. As both the $J_{spin}/J_{orb}$ value and the mass ratio exceed the instability boundary, it should be a contact binary merger candidate. Future observations should be carried out to continuously monitor this system in order to observe its merging process.

\section{Data availability}
Table A3, Figs. A3 and A4 are only available in electronic form at the CDS via anonymous ftp to cdsarc.u-strasbg.fr (130.79.128.5) or via http://cdsweb.u-strasbg.fr/cgi-bin/qcat?J/A+A/.

\begin{acknowledgements}
This work is supported by National Natural Science Foundation of China (NSFC) (No. 12273018), and the Joint Research Fund in Astronomy (No. U1931103) under cooperative agreement between NSFC and Chinese Academy of Sciences (CAS), and by the Qilu Young Researcher Project of Shandong University, and by the Young Data Scientist Project of the National Astronomical Data Center, and by the Cultivation Project for LAMOST Scientific Payoff and Research Achievement of CAMS-CAS. The calculations in this work were carried out at Supercomputing Center of Shandong University, Weihai.\\

We acknowledge the support of the staff of the Xinglong 2.16m, 85cm
telescopes, Lijiang 2.4m Telescope and WHOT. This work was partially supported by the Open Project Program of the Key Laboratory of Optical Astronomy, National Astronomical Observatories, Chinese Academy of Sciences.\\

The spectral data were provided by Guoshoujing Telescope (the Large Sky Area Multi-Object Fiber Spectroscopic Telescope LAMOST), which is a National Major Scientific Project built by the Chinese Academy of Sciences. Funding for the project has been provided by the National Development and Reform Commission. LAMOST is operated and managed by the National Astronomical Observatories, Chinese Academy of Sciences.\\

This paper makes use of data from the DR1 of the WASP data \citep{2010A&A...520L..10B} as provided by the WASP consortium,
and the computing and storage facilities at the CERIT Scientific Cloud, reg. no. CZ.1.05/3.2.00/08.0144 which is operated by Masaryk University, Czech Republic.\\

This work includes data collected by the TESS mission. Funding for the TESS mission is provided by NASA Science Mission Directorate. We acknowledge the TESS team for its support of this work.\\

This work has made use of data from the European Space Agency (ESA) mission
{\it Gaia} (\url{https://www.cosmos.esa.int/gaia}), processed by the {\it Gaia}
Data Processing and Analysis Consortium (DPAC,
\url{https://www.cosmos.esa.int/web/gaia/dpac/consortium}). Funding for the DPAC
has been provided by national institutions, in particular the institutions
participating in the {\it Gaia} Multilateral Agreement.\\

\end{acknowledgements}

%
%

\bibliography{sample631}
\bibliographystyle{aa}

\begin{appendix}

\renewcommand\arraystretch{1.3}
\setcounter{table}{0}
\renewcommand{\thetable}{A\arabic{table}}

\section{Considering whether there is a bright companion around TYC 3801-1529-1}\label{sec:bright companion}
\textit{Gaia} DR3 \citep{2016A&A...595A...1G,2023A&A...674A...1G} identified that there is a bright companion whose G band magnitude is 12$^m$.775 around TYC 3801-1529-1 within 0.5$^{\prime\prime}$, the brightness of the companion is similar to that of TYC 3801-1529-1 (G band magnitude of TYC 3801-1529-1 is 12$^m$.758). Such light contamination can squeeze the light curve amplitude and therefore not only the eclipses, but also the out-of-eclipse variation (i.e. mainly the ellipsoidal effect which is the basis of the mass ratio determination) is squeezed, leading to a too small (i.e. falsified) mass ratio. However, based on our carefully analysis, we found that this bright companion does not exist. There are three reasons. First, from the CCF curves shown in Fig. \ref{Fig3A}, we can see only one obvious peak from -800 km/s to 800 km/s. Secondly, observation images from WHOT, XL85, and WH50, the stellar image of TYC 3801-1529-1 is a circle, and only one star's PSF can be obtained. Thirdly, if the bright companion is really existing, neither ZTF nor Pan-STARRS can resolve these two stars, the g band magnitude of the two surveys should be smaller than 12$^m$ (two stars with nearly identical magnitudes together result in a magnitude decrease of 0$^m$.75). However, the ZTF mean g band magnitude of TYC 3801-1529-1 is 12$^m$.537 \citep{2020ApJS..249...18C}, and the PanSTARRS mean g band magnitude is 12$^m$.508 \citep{2016arXiv161205560C}, which are similar to the G band magnitude. In addition, we tried to use the Lijiang 2.4m Telescope \citep{2019RAA....19..149W} to provide lucky imaging evidence to disprove the existence of a close companion. The PI 2K$\times$2K CCD was equipped on this telescope, and the pixel scale is 0.283$^{\prime\prime}$, resulting a field of view of 9.6$^{\prime}$x9.6$^{\prime}$. TYC 3801-1529-1 was observed on September 20, 2024 using the R band filter. 57 effective images with exposure time of 0.1s were obtained, and 100 effective images with exposure time of 1s were obtained. During the observations, seeing is about 1.19$^{\prime\prime}$. In order to check whether there is a close companion star around our target, we created a 25$\times$25 pixels image centered on our selected stars, along with a flux distribution map, and the flux distribution fitting using a 2D Gaussian function, they are shown in Fig. \ref{Fig4A}. Based on the above content, we created an animation which can be seen from Fig. \ref{Fig4A} in the online version of this article or from China VO (DOI: 10.12149/101494)\footnote{https://nadc.china-vo.org/res/r101494/}. Three stars were selected for comparison: TYC 3801-1734-1 (G=12.177 mag), Gaia DR3 1034289323867544832 (G=13.086 mag), and Gaia DR3 1031285561179312000 (G=14.079 mag). Our target is illustrated in the first line of the animation, TYC 3801-1734-1 is shown in the second line, Gaia DR3 1034289323867544832 is displayed in the third line, while Gaia DR3 1031285561179312000 is presented in the fourth line. From the animation, we can see that the star images and the distribution of their flux are almost consistent. Then, we calculated the full width at half maximum (FWHM) of the 2D Gaussian fitting of the flux distribution, $(x^2+y^2)^{0.5}$ of FWHM is 2.537$\pm0.343$ for our target, 2.543$\pm0.398$ for TYC 3801-1734-1, 2.561$\pm0.445$ for Gaia DR3 1034289323867544832, and 2.442$\pm0.392$ for Gaia DR3 1031285561179312000. The values of FWHM of the four stars are consistent with each other. Based on these results, we believe that we can rule out the existence of the bright close companion.

\begin{table}[htbp]
\scriptsize
\begin{flushleft}
\caption{ Observation log of TYC 3801-1529-1.} \label{tab:observation}
\resizebox{\columnwidth}{!}{
\begin{tabular}{cccccc}
\hline
\multicolumn{6}{c}{Photometric observations}\\ \hline
Time (UT)    & Filter         & Exposure (s) & Mean errors (mag)       & Type   &Telescope   \\
\hline
Feb 10, 2021 & V              & 60           & 0.005                   & Light curve & WHOT  \\
Feb 11, 2021 & R$_c$/I$_c$    & 20/15        & 0.008/0.009             & Light curve & WHOT  \\
Feb 12, 2021 & B              & 90           & 0.007                   & Light curve & WHOT  \\
Feb 22, 2021 & R$_c$/I$_c$    & 40/40        & 0.009/0.009             & Light curve & XL85  \\
Jan 20, 2022 & B/V/R$_c$/I$_c$& 90/35/15/15  & 0.005/0.009/0.005/0.005 & Light curve & WHOT   \\
Jan 28, 2023 & B/V/R$_c$/I$_c$&100/50/35/35  & 0.005/0.004/0.005/0.004 & Light curve & XL85   \\
Dec 04, 2023 & R$_c$          & 40           & 0.003                   & Minimum     & XL85   \\
Mar 01, 2024 & B/V/R$_c$/I$_c$& 90/35/15/15  & 0.005/0.009/0.005/0.005 & Minimum     & WH50   \\
Mar 02, 2024 & B/V/R$_c$/I$_c$&100/50/25/20  & 0.009/0.008/0.007/0.007 & Minimum     & WHOT   \\
Mar 06, 2024 & B/V/R$_c$/I$_c$&100/50/25/20  & 0.008/0.009/0.007/0.007 & Light curve & WHOT   \\
Mar 08, 2024 & B/V/R$_c$/I$_c$&100/50/25/20  & 0.004/0.005/0.007/0.004 & Light curve & WHOT   \\\hline
\multicolumn{6}{c}{Spectroscopic observations}\\ \hline
Time (UT)    &  Exposure (s) & Number of Frames & Signal Noise Ratio$^*$   &Telescope   \\
\hline
Feb 22, 2021 &  1800 & 7 & $35-45$   & XL216   \\
Jan 05, 2022 &  1800 & 11& $45-55$   & XL216   \\
Jan 27, 2022 &  1800 & 4 & $40-50$   & XL216   \\
\hline
\end{tabular}}
\end{flushleft}
$^*$ The Signal Noise Ratio is around the wavelength of 5500 {\AA}.
\end{table}

\begin{table}[htbp]
\scriptsize
\begin{center}
\caption{ Radial velocities of TYC 3801-1529-1.} 
\begin{tabular}{cccc}
\hline
Time (HJD)    & Phase         & RV$_1$ (km/s) & Errors   \\
\hline
2459267.99633  & 0.076  & -8.3   & 5.0   \\
2459268.01725  & 0.133  & -8.1   & 5.1   \\
2459268.03818  & 0.190  & -8.5   & 5.3   \\
2459268.10086  & 0.361  & -4.3   & 6.1   \\
2459268.16747  & 0.543  & 11.2   & 5.7   \\
2459268.18839  & 0.601  & 11.8   & 4.7   \\
2459268.20933  & 0.658  & 13.6   & 7.2   \\
2459585.11888  & 0.721  & 16.9   & 4.2   \\
2459585.13979  & 0.778  & 16.5   & 4.8   \\
2459585.16776  & 0.855  & 6.9    & 4.4   \\
2459585.18867  & 0.912  & 7.3    & 4.6   \\
2459585.28476  & 0.174  & -11.3  & 4.2   \\
2459585.30569  & 0.231  & -18.3  & 4.0   \\
2459585.33350  & 0.307  & -15.3  & 4.5   \\
2459585.35442  & 0.365  & -15.2  & 4.5   \\
2459585.38194  & 0.440  & -6.8   & 4.6   \\
2459585.40286  & 0.497  & -2.4   & 4.4   \\
2459585.43047  & 0.572  & -1.4   & 4.4   \\
2459607.08769  & 0.758  & 6.8    & 5.2   \\
2459607.11918  & 0.844  & 5.5    & 4.9   \\
2459607.26099  & 0.232  & -5.1   & 5.6   \\
2459607.29278  & 0.319  & -5.0   & 4.9   \\
\hline
\end{tabular}
\label{tab:RV}
\end{center}
Note.$-$The phase was calculated by the equation, HJD=2459271.261869+0$^d$.36591971$\times$E.
\end{table}

\setcounter{figure}{0}
\renewcommand{\thefigure}{A\arabic{figure}}
\begin{figure}[htbp]
\centering
\includegraphics[scale=0.35]{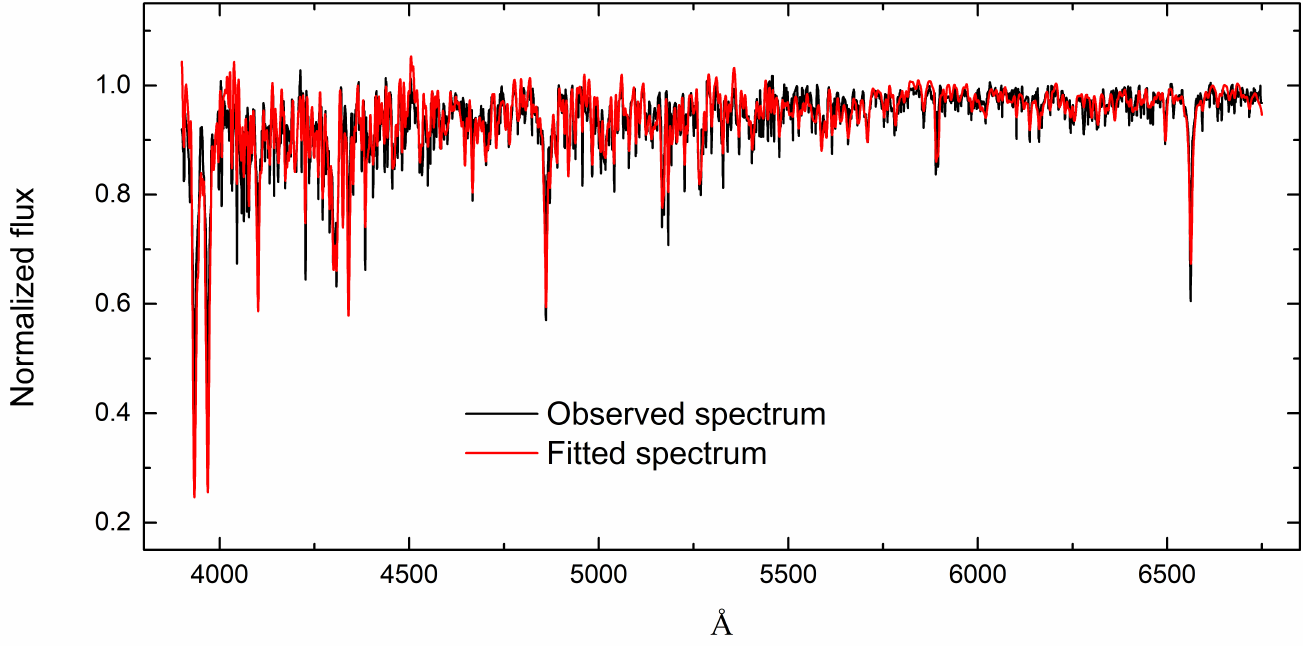}
\caption{Normalised spectrum at phase=0.497 and fitted spectrum of TYC 3801-1529-1.}
\label{Fig1A}
\end{figure}

\renewcommand\arraystretch{1.1}
\begin{table}[htbp]
\scriptsize
\begin{center}
\caption{ Eclipse timings of TYC 3801-1529-1.} 
\begin{tabular}{ccccc}
\hline
BJD$_{TDB}$    & Error   &  E        &  O-C      &   Source     \\\hline
2454427.66935  & 0.00254 &  -13196   &  0.08616  &   SuperWASP  \\
2454436.62525  & 0.00210 &  -13171.5 &  0.07702  &   SuperWASP  \\
2454437.72049  & 0.00128 &  -13168.5 &  0.07451  &   SuperWASP  \\
2454438.63216  & 0.00181 &  -13166   &  0.07138  &   SuperWASP  \\
2454496.45589  & 0.00451 &  -13008   &  0.07978  &   SuperWASP  \\
2454501.57944  & 0.00096 &  -12994   &  0.08046  &   SuperWASP  \\
2454502.48347  & 0.00133 &  -12991.5 &  0.06969  &   SuperWASP  \\
2454523.35892  & 0.00103 &  -12934.5 &  0.08771  &   SuperWASP  \\
2454524.44560  & 0.00104 &  -12931.5 &  0.07664  &   SuperWASP  \\
2454525.37062  & 0.00137 &  -12929   &  0.08686  &   SuperWASP  \\
2454526.45986  & 0.00124 &  -12926   &  0.07834  &   SuperWASP  \\
2454527.38443  & 0.00095 &  -12923.5 &  0.08811  &   SuperWASP  \\
2454527.55667  & 0.00129 &  -12923   &  0.07739  &   SuperWASP  \\
2454530.49015  & 0.00151 &  -12915   &  0.08351  &   SuperWASP  \\
2454532.49583  & 0.00119 &  -12909.5 &  0.07663  &   SuperWASP  \\
\hline
\end{tabular}
\label{tab:ET}
\end{center}
Note.$-$This table is available in its entirety in machine-readable form in the online journal. A portion is shown here for guidance regarding its form and content.
\end{table}

\begin{figure*}
\centering
\includegraphics[width=0.32\textwidth]{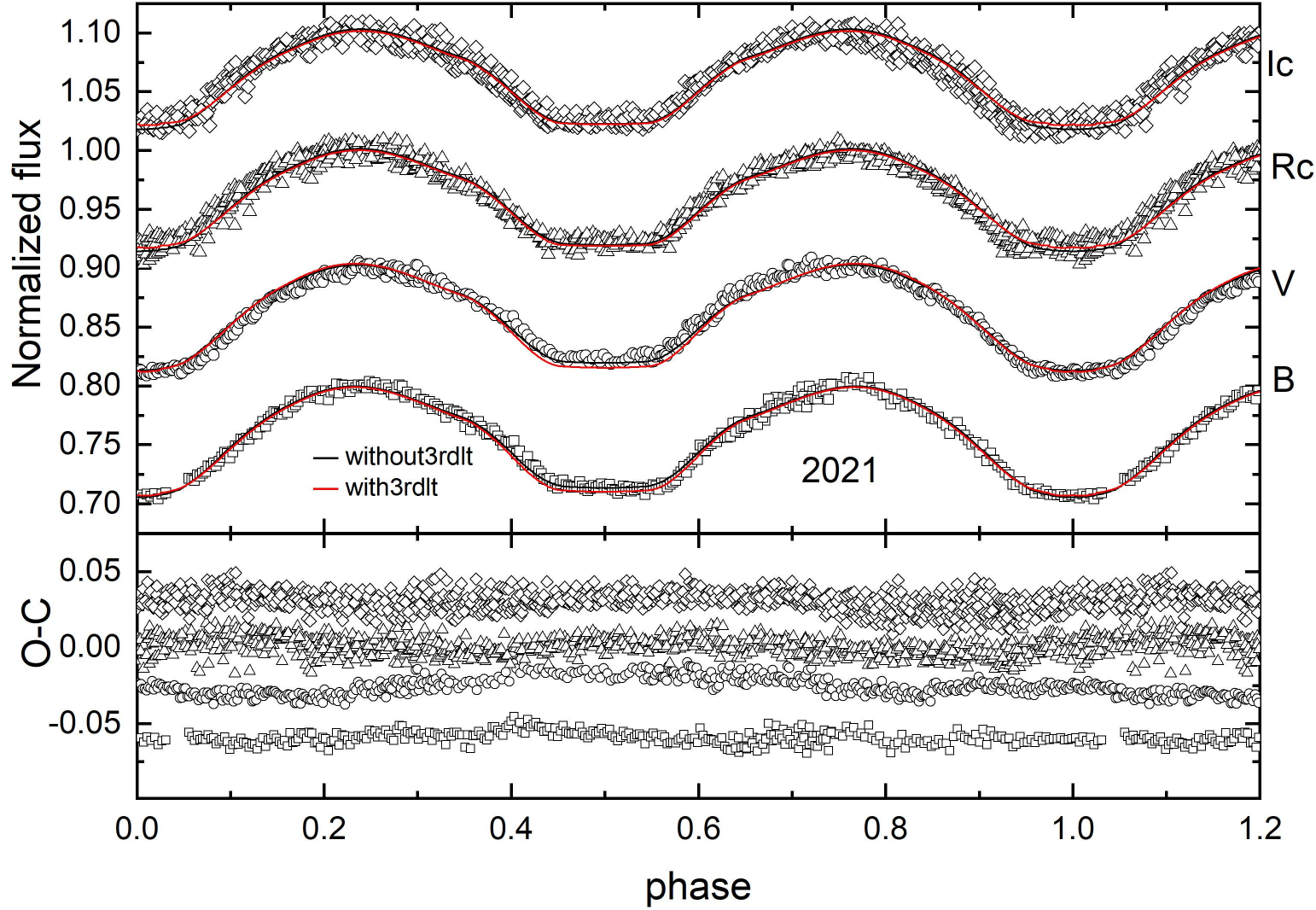}
\includegraphics[width=0.32\textwidth]{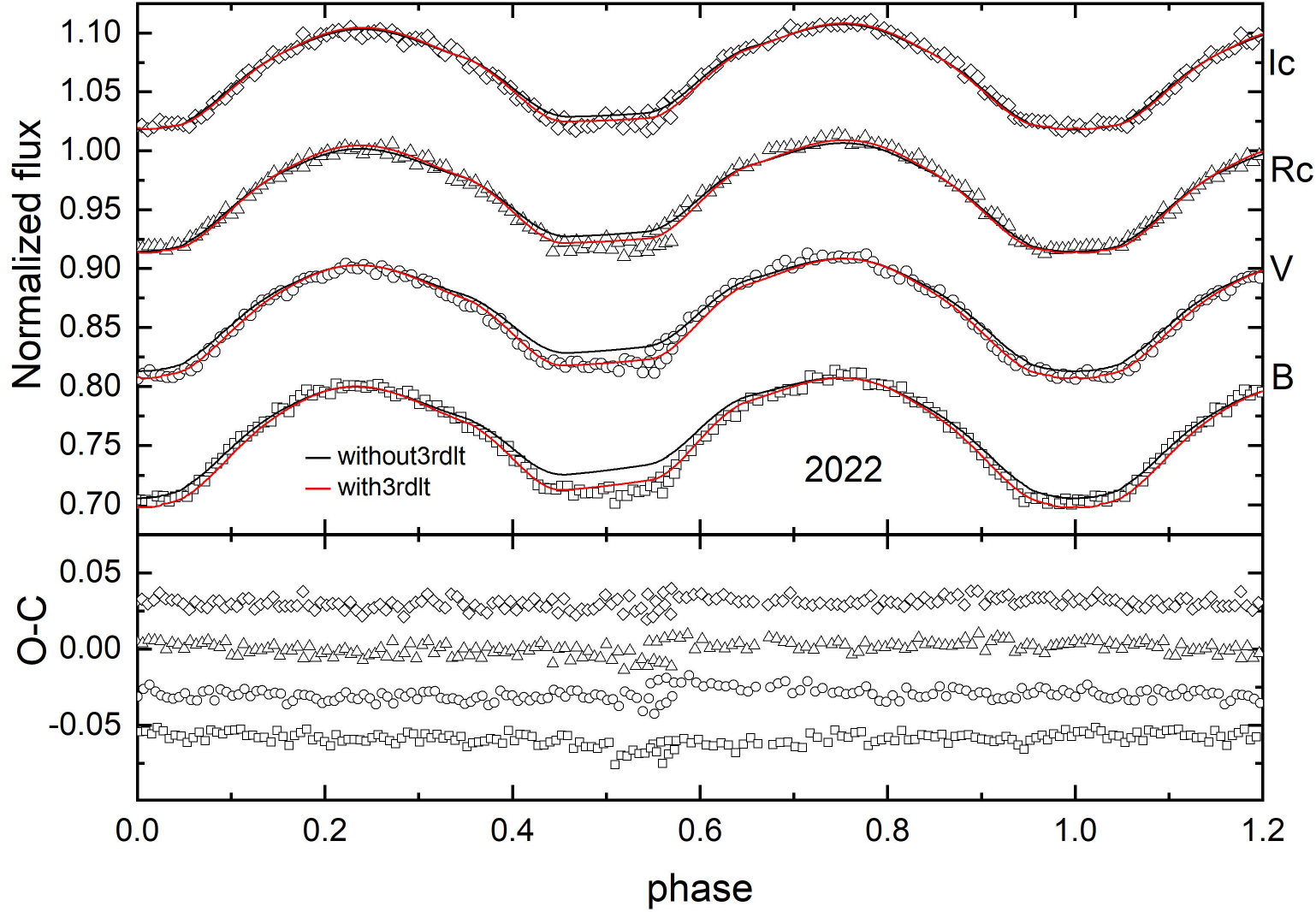}
\includegraphics[width=0.32\textwidth]{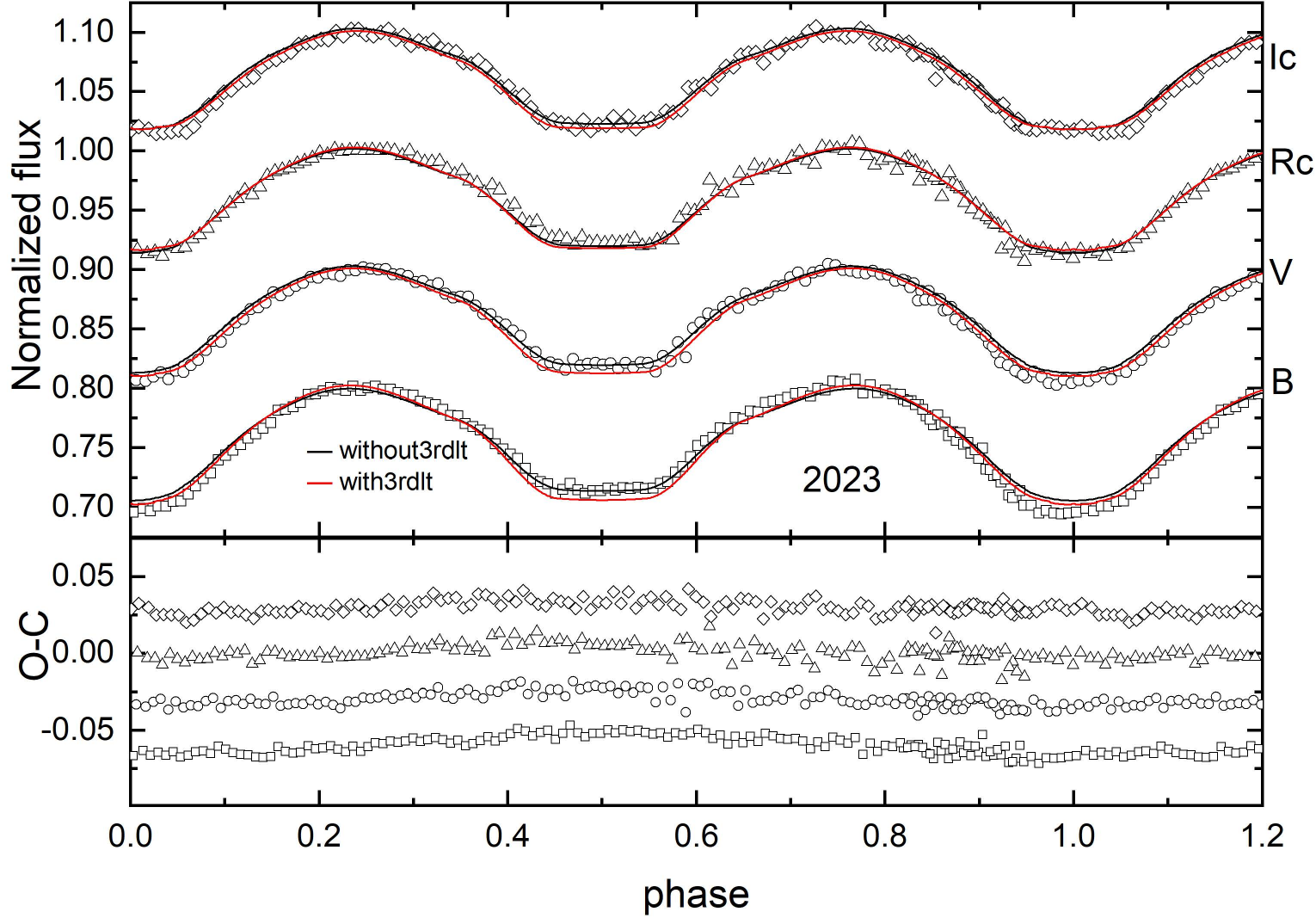}
\includegraphics[width=0.32\textwidth]{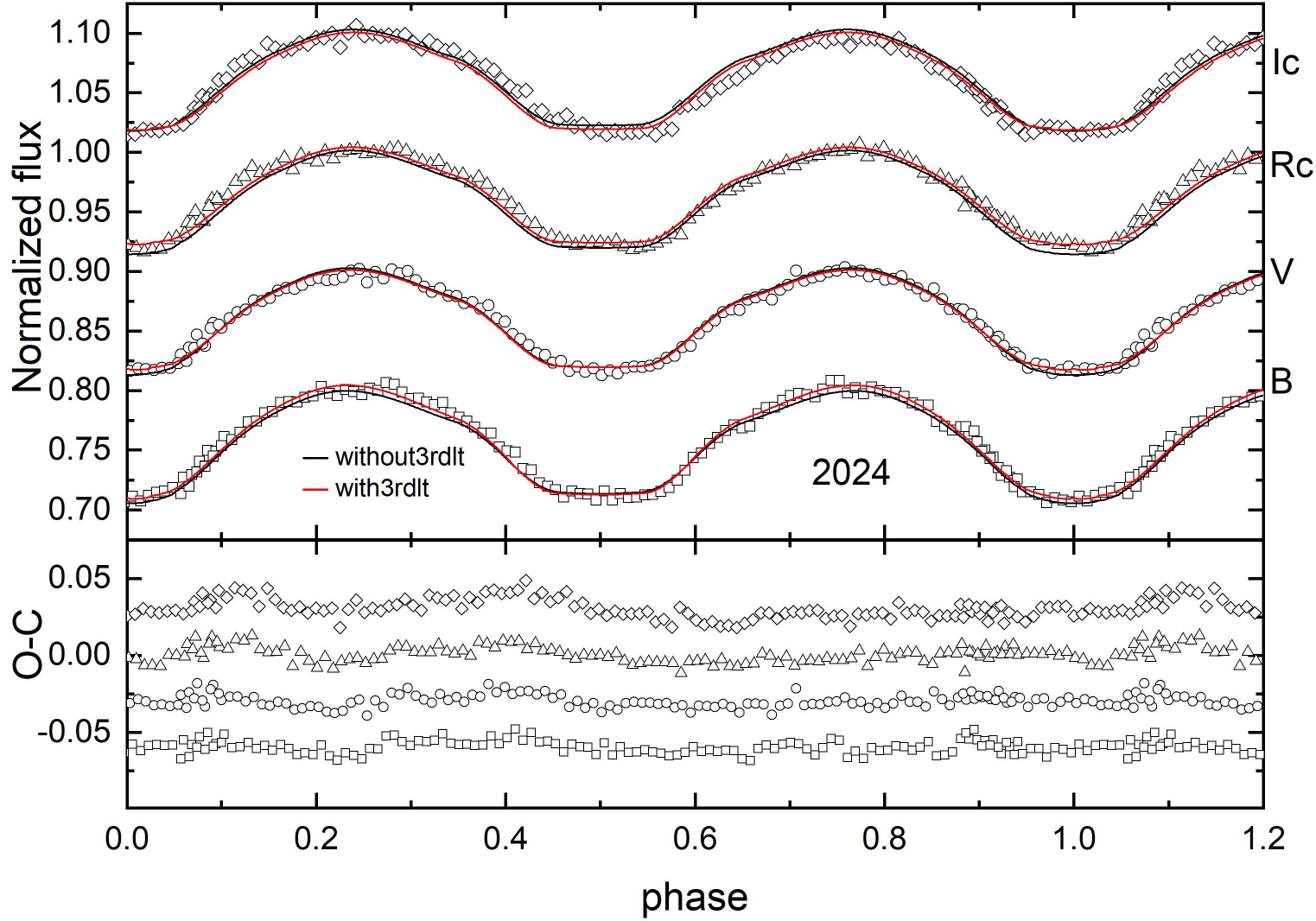}
\includegraphics[width=0.32\textwidth]{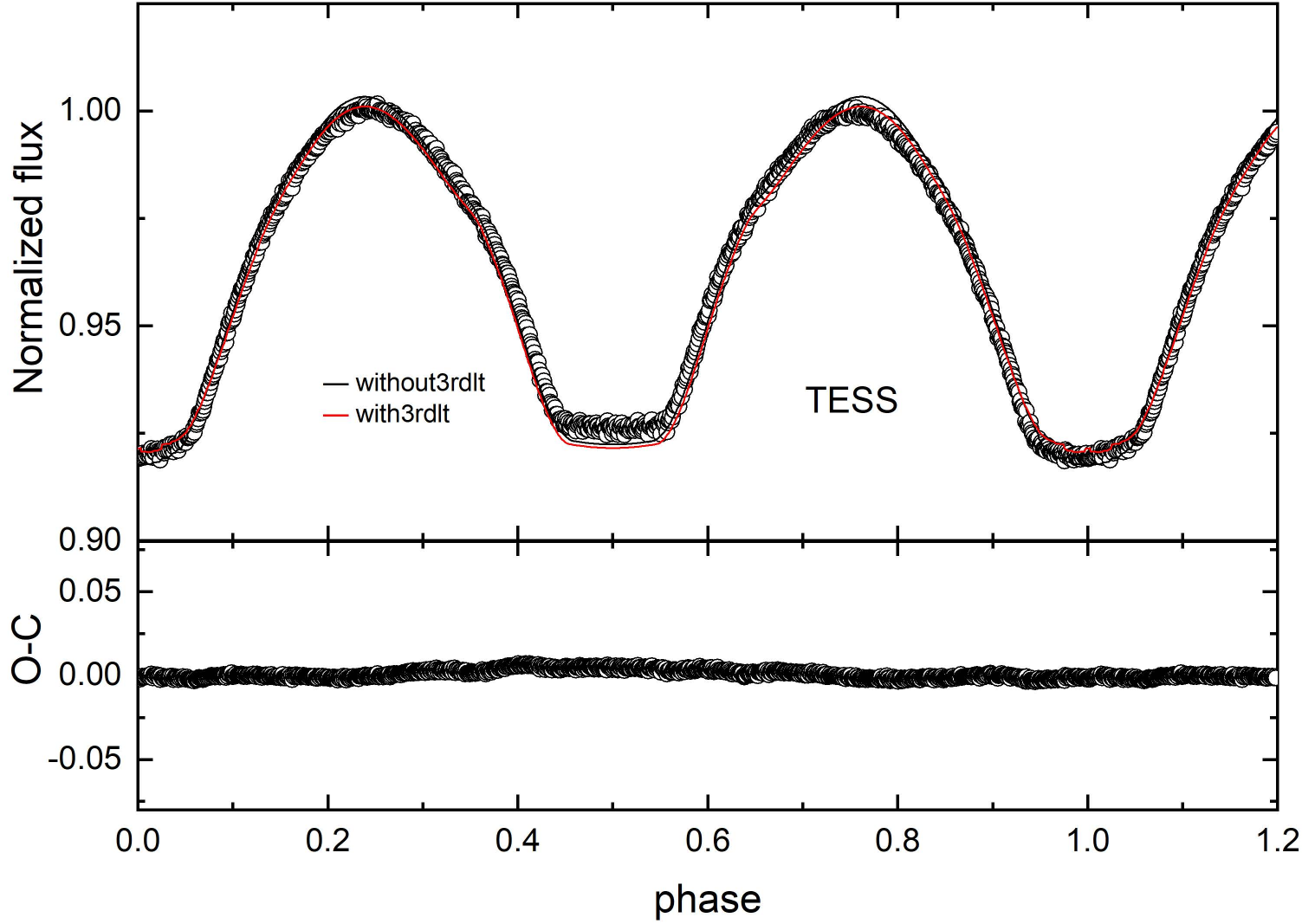}
\caption{Comparison between the theoretical light curves and the observed ones. Black lines show the theoretical light curves without third light, while the red ones represent those with third light.}
\label{Fig2A}
\end{figure*}

\begin{figure}
\centering
\includegraphics[width=0.35\textwidth]{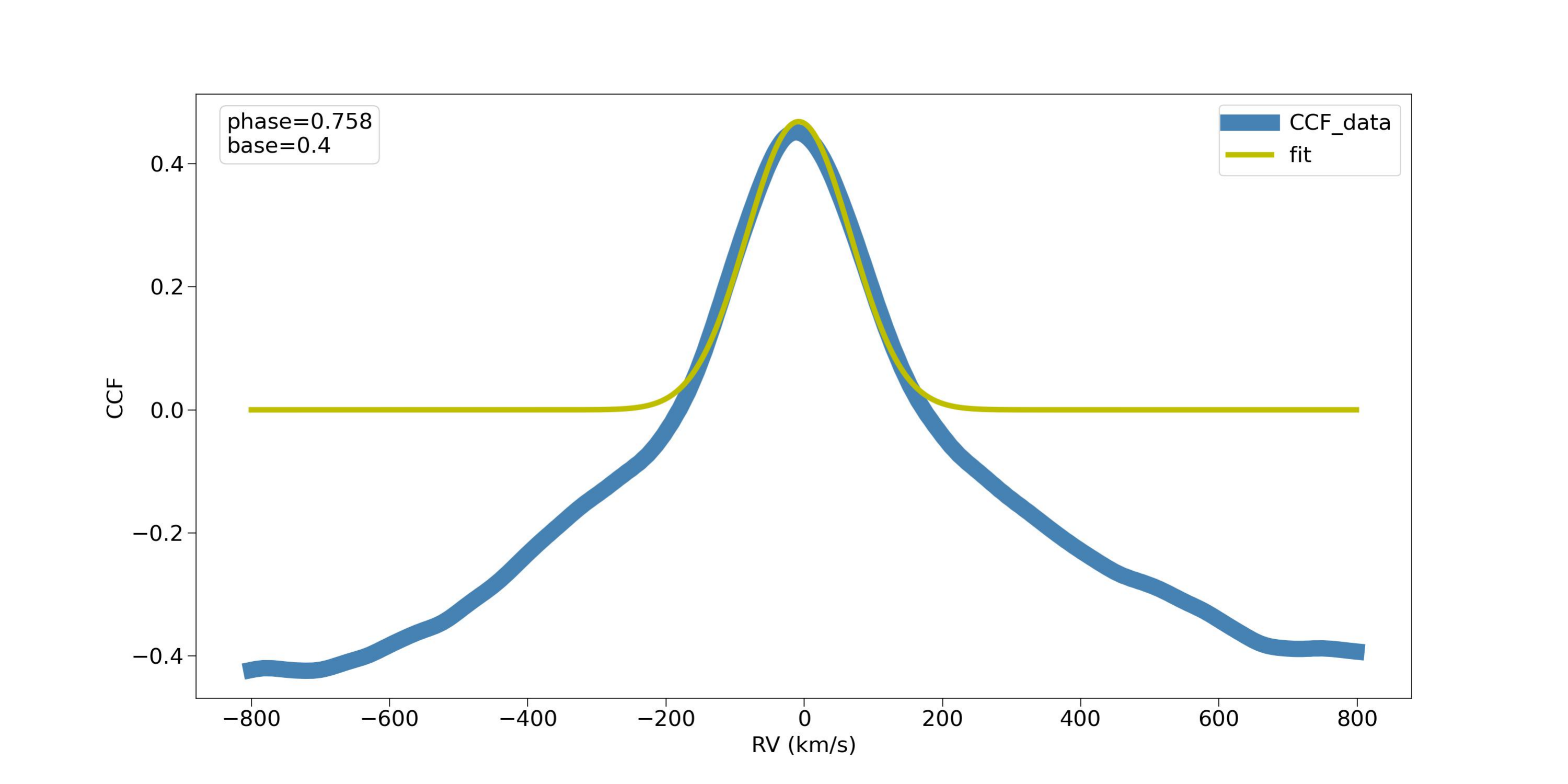}
\caption{CCF curve of TYC 3801-1529-1. Phase=0.758 is shown here for example. The complete CCF curves can be found in the online version of this article or from China VO (DOI: 10.12149/101494).}
\label{Fig3A}
\end{figure}

\begin{figure}
\centering
\includegraphics[width=0.35\textwidth]{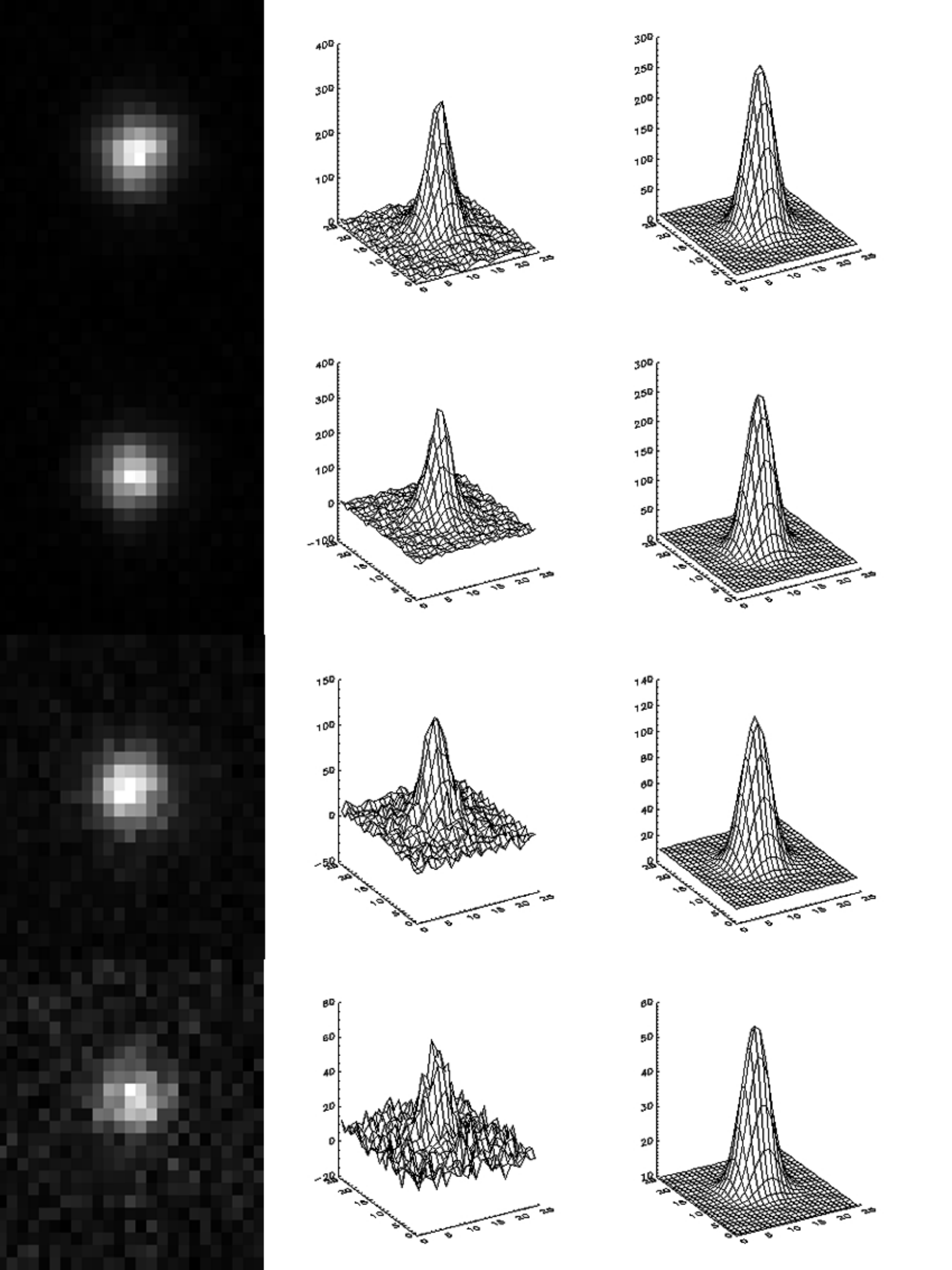}
\caption{25$\times$25 pixel image centered on our selected stars, along with a flux distribution map, and the flux distribution fitting using a 2D Gaussian function, we created an animation which can be seen in the online version of this article or from China VO (DOI: 10.12149/101494).}
\label{Fig4A}
\end{figure}

\end{appendix}

\end{document}